\documentclass{article}

\usepackage{arxiv}

\usepackage[utf8]{inputenc} % allow utf-8 input
\usepackage[T1]{fontenc}    % use 8-bit T1 fonts
\usepackage{hyperref}       % hyperlinks
\usepackage{url}            % simple URL typesetting
\usepackage{booktabs}       % professional-quality tables
\usepackage{amsfonts}       % blackboard math symbols
\usepackage{nicefrac}       % compact symbols for 1/2, etc.
\usepackage{microtype}      % microtypography
\usepackage{lipsum}		% Can be removed after putting your text content
\usepackage{graphicx}
\usepackage{natbib}
\usepackage{doi}

\hypersetup{
  colorlinks   = true, %Colours links instead of ugly boxes
  urlcolor     = blue, %Colour for external hyperlinks
  linkcolor    = red, %Colour of internal links
  citecolor   = green %Colour of citations
}
\setcitestyle{numbers,open={[},close={]}}
\usepackage{subcaption}
\usepackage{adjustbox}
\usepackage{multirow}
\title{Fibroglandular Tissue Segmentation in Breast MRI using Vision Transformers - A multi-institutional evaluation}

\author{
        \href{https://orcid.org/0000-0002-7413-2570}{\includegraphics[scale=0.06]{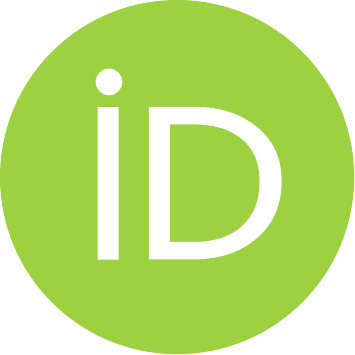}\hspace{1mm}Gustav Müller-Franzes} 
	\And
        \href{https://0009-0006-8684-9161}{\includegraphics[scale=0.06]{orcid.pdf}\hspace{1mm}Fritz Müller-Franzes} 
	\And
         \href{https://orcid.org/0000-0002-4898-4294}{\includegraphics[scale=0.06]{orcid.pdf}\hspace{1mm}Luisa Huck} 
        \And
         \href{https://orcid.org/0000-0002-8431-6752}{\includegraphics[scale=0.06]{orcid.pdf}\hspace{1mm}Vanessa Raaff} 
	\And
        \href{https://orcid.org/0009-0000-0434-7093}{\includegraphics[scale=0.06]{orcid.pdf}\hspace{1mm}Eva Kemmer} 
	\And
        \href{https://orcid.org/0000-0001-5089-3589}{\includegraphics[scale=0.06]{orcid.pdf}\hspace{1mm}Firas Khader} 
	\And
         \href{https://orcid.org/0000-0003-1015-7733}{\includegraphics[scale=0.06]{orcid.pdf}\hspace{1mm}Soroosh Tayebi Arasteh} 
         \And
         \href{https://orcid.org/0000-0003-3167-7121}{\includegraphics[scale=0.06]{orcid.pdf}\hspace{1mm}Teresa Nolte}
        \And
	\href{https://orcid.org/0000-0002-3730-5348}{\includegraphics[scale=0.06]{orcid.pdf}\hspace{1mm}Jakob Nikolas Kather} 
         \And
         \href{https://orcid.org/0000-0002-5267-9962}{\includegraphics[scale=0.06]{orcid.pdf}\hspace{1mm}Sven Nebelung} 
	\And
         \href{https://orcid.org/0000-0001-8696-2363}{\includegraphics[scale=0.06]{orcid.pdf}\hspace{1mm}Christiane Kuhl} 
        \And
	\href{https://orcid.org/0000-0002-9605-0728}{\includegraphics[scale=0.06]{orcid.pdf}\hspace{1mm}Daniel Truhn} 
}

\date{}

\renewcommand{\undertitle}{}
\hypersetup{
pdftitle={A template for the arxiv style},
pdfsubject={cs.CV},
pdfauthor={},
pdfkeywords={},
}

\begin{document}
\maketitle

\begin{abstract}
Accurate and automatic segmentation of fibroglandular tissue in breast MRI screening is essential for the quantification of breast density and background parenchymal enhancement. In this retrospective study, we developed and evaluated a transformer-based neural network for breast segmentation (TraBS)  in multi-institutional MRI data, and compared its performance to the well established convolutional neural network nnUNet.

TraBS and nnUNet were trained and tested on 200 internal and 40 external breast MRI examinations using manual segmentations generated by experienced human readers. Segmentation performance was assessed in terms of the Dice score and the average symmetric surface distance.

The Dice score for nnUNet was lower than for TraBS on the internal testset (0.909±0.069 versus 0.916±0.067, P<0.001) and on the external testset (0.824±0.144 versus 0.864±0.081,  P=0.004). Moreover, the average symmetric surface distance was higher (=worse) for nnUNet than for TraBS on the internal (0.657±2.856 versus 0.548±2.195, P=0.001) and on the external testset (0.727±0.620 versus 0.584±0.413, P=0.03).

Our study demonstrates that transformer-based networks improve the quality of fibroglandular tissue segmentation in breast MRI compared to convolutional-based models like nnUNet. These findings might help to enhance the accuracy of breast density and parenchymal enhancement quantification in breast MRI screening.

\end{abstract}

% keywords can be removed

\section{Introduction}

Breast cancer is the most frequent type of cancer in the female population, and represents the second leading cause of death in the United States \cite{Siegel2022-kj} among women. New guidelines for breast cancer screening recommend the use of MRI for women with dense breast tissue, whereas X-ray mammography was previously the primary imaging modality for breast cancer screening \cite{mann2022, Oeffinger2015-yz}. Deep learning-based tools for the assessment of breast density on mammography have already been developed \cite{Magni2022-av}, yet a consistent and reliable automated assessment of breast density - as the ratio of fibroglandular tissue (FGT) to the breast volume - on MRI examinations is still lacking. Besides breast density, the enhancement of fibroglandular tissue (BPE) has also emerged as a promising marker for the early detection of breast cancer \cite{sindi2019, thompson2019}, however, reliable automated assessment of BPE is also lacking. The development of a machine learning algorithm capable of segmenting the fibgroglandular tissue (FGT) is an important first step towards an automatic quantification of breast density and BPE in breast MRI examinations. 
Several research studies have investigated this problem by training convolutional neural networks (CNNs) on manually segmented breast MRI examinations and evaluating their performance on single-center test sets \cite{dalmis2017, huo2021, zhang2019}. The high level of agreement between human- and machine-generated segmentation maps in all of these publications demonstrates the potential of CNNs. However, there is an important impediment to the widespread introduction of such algorithms: MRI examinations are not standardized. Different clinical centers use diverse MRI protocols and sequences for the diagnosis of breast cancer. None of the studies we found tested their CNN architecture on independent data that did not belong to the institution where the algorithms were developed. 
Transformer-based models have proven to be more robust, generalizable, and attack-proof than CNNs in other applications of medical image analysis \cite{ghaffarilaleh2022a, tang2022}. They have achieved state-of-the-art results for natural language processing \cite{devlin2019, radforda}, mainly because of their capability to handle long-term dependencies and self-supervised pre-training for downstream tasks.
Therefore, we aimed to develop and test a robust and accurate segmentation method based on the transformer architecture that could generalize well to multi-institutional data.
We compare our model against the current state-of-the-art CNN-based model on both internal and external breast MRI datasets from Duke University \cite{saha2022}. Our hypotheses were that the new transformer-based model outperforms the current state of the art and that it generalizes better to external data.

\section{Material and Methods}

\subsection*{Ethics Statement}
Local institutional review board approval was obtained (EK028/19). 
\subsection*{Datasets}
In this retrospective study, two breast MRI datasets were used which we will refer to as “UKA”, “DUKE”. First, UKA was collected between 2010 and 2019 at the University Hospital Aachen, Germany \cite{muller-franzes2023}. UKA comprises a total of 9751 breast MRI examinations of 5086 women. Among this set, a total of 200 examinations from 200 women were chosen, comprising 104 carcinomas and 55 fibroadenomas. Dynamic Contrast Enhancement (DCE)-MRI studies of the breast had been performed according to a standardized protocol \cite{kuhl2017} on a 1.5-T system (Achieva and Ingenia; Philips Medical Systems) by using a double-breast four-element surface coil (Invivo) with two paddles being used to immobilize the breast in the craniocaudal direction (Noras). See Table \ref{tab:1} for a detailed description of the acquisition parameters.
Second, DUKE was collected between 2000 and 2014 at the Duke Hospital, USA, and is publicly available \cite{saha2022}. All 922 cases have biopsy-confirmed invasive breast cancer and were acquired with either a 1.5 Tesla, 2.9 Tesla, or 3.0 Tesla scanner from General Electric or Siemens.  The MRI protocol consisted of a T1-weighted fat-suppressed sequence (one pre-contrast, and four post-contrast scans) and a non-fat-suppressed T1-weighted sequence. For evaluation, 40 cases were randomly selected and manually segmented as detailed below.
Ground Truth Segmentation of Fibroglandular Tissue
Both the whole breast volume and the fibroglandular tissue were segmented by F.M. and E.K. using the software ITK-SNAP \cite{yushkevich2006}, quality controlled by L.H. and V.R. with six and three years of experience in breast MRI respectively, and corrected if necessary. Segmentation masks were generated for the UKA subset of 200 MRI examinations and 40 randomly sampled cases of DUKE, respectively. The breast outline was defined as the tissue volume located anterior to the pectoralis muscle. Sample manual segmentations are shown in Supplemental Figures \ref{fig:figS1} and \ref{fig:figS2}.
\begin{table}[]
\caption{Image acquisition parameters.  NA = Not Available, GR = Gradient Echo, SE = Spin Echo, DCE = Dynamic Contrast Enhancement  }
\label{tab:1}
\centering
\begin{tabular}{@{}lllll@{}}
\toprule
                       & \multicolumn{2}{l}{UKA}       & \multicolumn{2}{l}{DUKE}        \\ \midrule
Country                & \multicolumn{2}{l}{Germany}   & \multicolumn{2}{l}{USA}         \\
Orientation            & \multicolumn{2}{l}{Axial}     & \multicolumn{2}{l}{Axial}       \\
Scanner                & \multicolumn{2}{l}{Philips}   & \multicolumn{2}{l}{GE, Siemens} \\
Field   Strength [T]   & \multicolumn{2}{l}{1.5}       & \multicolumn{2}{l}{1.5, 3}      \\
Acquisition            & \multicolumn{2}{l}{2010-2019} & \multicolumn{2}{l}{2000-2014}   \\
                       & DCE           & T2            & DCE             & T1            \\
Acquisition   Type     & 2D            & 2D            & 3D              & 2D, 3D        \\
Echo   Type            & GR            & SE            & GR              & SE, GR        \\
Fat   Suppression      & No            & No            & Yes             & No            \\
TR   [ms] & 264±22  & 4008±201 & GR:5.0±0.7 & \begin{tabular}[c]{@{}l@{}}2D-SE:605±66\\ 2D-RM:5090±191\\ 3D-GR:6.7±0.7\end{tabular} \\
TE   [ms] & 4.6±0.1 & 110      & GR:2.1±0.5 & \begin{tabular}[c]{@{}l@{}}2D-SE:9.9±1.2\\ 3D-GR:3.5±0.8\end{tabular}                 \\
Flip   angle [°]       & 90±3          & 90            & 10±0.5          & 97±44         \\
Slice   Thickness [mm] & 3.1±0.2       & 3.1±0.2       & 1.2±0.3         & 2.3±0.7       \\
Number   of slices     & 28 ± 2        & 33±2          & 170±23          & 80±51         \\
Matrix   X             & 520±22        & 535±39        & 487±43          & 438±103       \\
Matrix   Y             & 520±22        & 535±39        & 487±43          & 438±103       \\
Field   of View X [mm] & 333±26        & 332±26        & 347±25          & 345±25        \\
Field   of View Y [mm] & 333±26        & 332±26        & 347±25          & 345±25        \\ \bottomrule
\end{tabular}
\end{table}
\subsection*{Data Processing Pipeline}
\begin{figure}
    \centering
    \includegraphics[width=\textwidth]{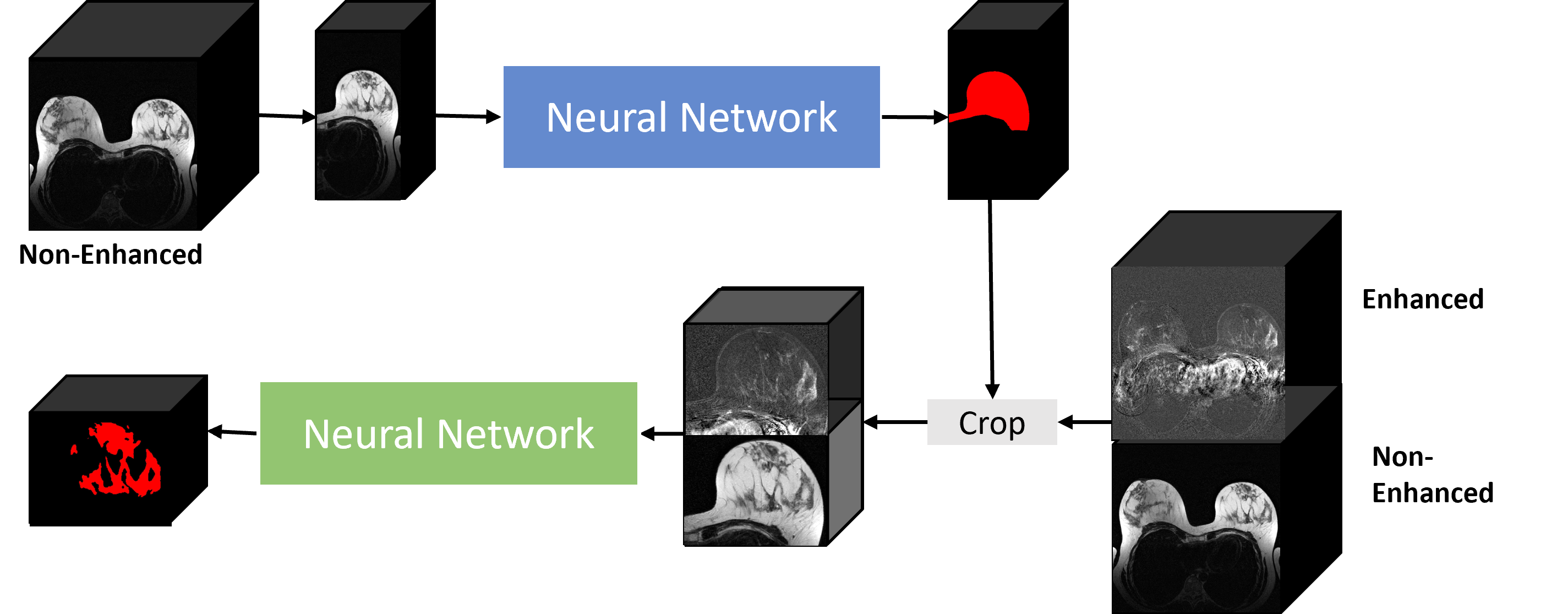}
    \caption{Illustration of the segmentation framework. Non-Enhanced T1- or T2-weighted images respectively (depending on the availability within the MRI protocol) were used to manually segment the breast. The manual breast segmentation is used to crop the subtraction image of the dynamic contrast enhanced (DCE) sequence and the pre-contrast image. Based on the cropped sequences, the neural network created a segmentation mask of the fibroglandular tissue.}
    \label{fig:1}
\end{figure}

The processing pipeline comprised two consecutive stages: The first stage performed the segmentation of the whole breast, while the second stage segmented the FGT only (Figure \ref{fig:1}). In both stages, the use of a neural network was possible, however, the manual (ground truth) segmentations were used in the first stage with the rationale that we want to compare the network architectures for FGT only. 
For the second step of the segmentation pipeline, the segmentation masks from the first step were used to create a crop of the left and right breast. The non-enhanced and the contrast-enhanced images were stacked along the channel dimension and both breast sides were subsequently fed into the neural network. The intensity distributions of all images were z-score normalized (mean=0, standard deviation = 1). The segmentation pipeline was implemented with PyTorch \cite{paszke2019} on a computer equipped with an NVIDIA GeForce RTX 3090.

\subsection*{Model Architecture}
\begin{figure}
    \centering
    \includegraphics[width=\textwidth]{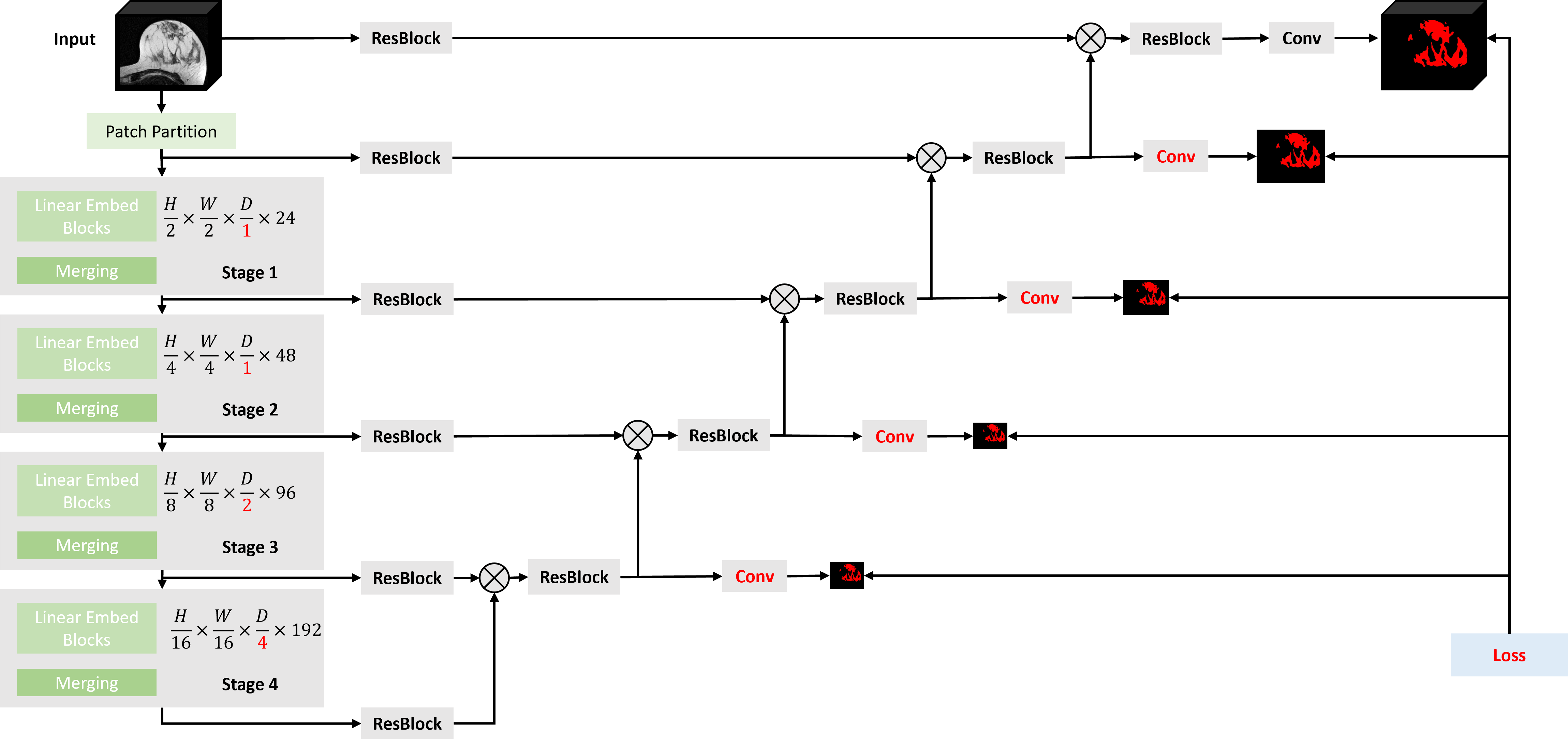}
    \caption{Illustration of the proposed TraBS model architecture. Non-isotropic kernels and strides were used in the first two stages in order for the depth to remain constant. Deep supervision was added for the lower-resolution layers. Changes from the original SwinUNETR have been marked in red. Please refer to the SwinUNETR publication\cite{hatamizadeh2022} for an in-depth explanation.}
    \label{fig:2}
\end{figure}

In the following, we refer to our new transformer-based model as TraBS (SwinTransformer for fibroglandular Breast tissue Segmentation). TraBS was built upon SwinUNETR \cite{hatamizadeh2022} with 2, 4, and 8 heads and 24, 48, 96, and 192 embedding features in stages 1 to 4. Inspired by the nnUNet to handle typically non-isotopic resolutions in MRI images, we replaced the uniform 2x2x2 patch sizes and 3x3x3 kernels in the two up-most layers with non-isotropic 1x2x2 patches and 1x3x3 kernels. In addition, 1x1x1 convolutions were added to supervise the deeper layers (Figure \ref{fig:2}).
To establish a baseline, we employed the state-of-the-art nnUNet \cite{isensee2021}. The model had two max-pooling layers with 1x2x2 strides and 1x3x3 kernels, followed by two max-pooling layers with 2x2x2 strides and 3x3x3 kernels, following a previous publication for FGT segmentation \cite{huo2021}.
Training and Testing of the Framework
The UKA subset was randomly divided into training and test sets using five-fold cross-validation. The training set within each fold was further subdivided into a dedicated training set (80\%) and a validation set (20\%). The training of the FGT segmentation models was performed for each of the five folds with the manual segmentation masks as ground truth. AdamW with a learning rate of 0.0001 was used to optimize the sum of DiceLoss and CrossEntropy, following previous recommendations for medical image segmentation \cite{ma2021}. Following the nnUNet implementations, the loss function was additionally calculated at the lower resolutions of the decoder path (Multi-Scale Supervision) in the TraBS model. Using early stopping, training of each model was halted as soon as the loss within the validation set did not decrease within 30 epochs. 
To increase the diversity of the training set and thus prevent overfitting, the following data augmentation operations from the TorchIO framework  \cite{perez-garcia2020} were applied: flipping, affine transformation, ghosting, Gaussian noise, blurring, bias field, and gamma augmentation. During training, a random region of 256x256x32 voxels within the left and right crops was selected. A sliding window of 256x256x32 voxels with an overlap of 50\% was used during inference. Random-flip along all axes was used as test-time augmentation. The source code is publicly available at \url{https://github.com/mueller-franzes/TraBS}. 

\subsection*{Statistical Analysis}
We used five-fold cross-validation on the internal UKA dataset to examine the performance of the models on unseen test data. For the external DUKE dataset, an ensemble of the five FGT segmentation models from the cross-validation training was applied. Majority voting was used to combine the five segmentation masks.  Segmentation performance was assessed by calculating the Dice similarity coefficient (DSC) \cite{zou2004} and Average Symmetric Surface Distance (ASSD) \cite{Nai2021-wo}. Breast density and the BPE are both clinically relevant metrics related to breast cancer risk \cite{sindi2019, thompson2019} and their quantitative assessment depends on the FGT segmentation. Therefore, we measured these two metrics both for the manual and the automated segmentations and calculated the Pearson correlation coefficients between manually and automatically derived metrics. Note, that the BPE was defined here as the percentage change of the FGT between the post- and pre-contrast image. Bootstrapping was employed to calculate confidence intervals and permutation testing was used to calculate p-values. Following the guidance of Amrhein et al. \cite{Amrhein2019-wr}, we did not employ thresholds for statistical significance to the p-values.
\subsection*{Data Availability}
The DUKE dataset analysed during the current study is available in The Cancer Imaging Archive, \url{https://doi.org/10.7937/TCIA.e3sv-re93}.
The UKA dataset analysed during the current study is available from the corresponding author on reasonable request.

\section{Results}

\subsection*{Patient Characteristics}
The study included only female patients, with a mean age of 56±10 years (range 19-91) and a mean weight of 75±27 kg for the UKA data and 53±11 years (range 22-90) years with a mean weight of 76±18kg for the DUKE data. 
\subsection{Transformer outperforms CNNs on the Internal Dataset}
\subsubsection*{Overall segmentation performance}
\begin{figure}
    \centering
    \includegraphics[width=\textwidth]{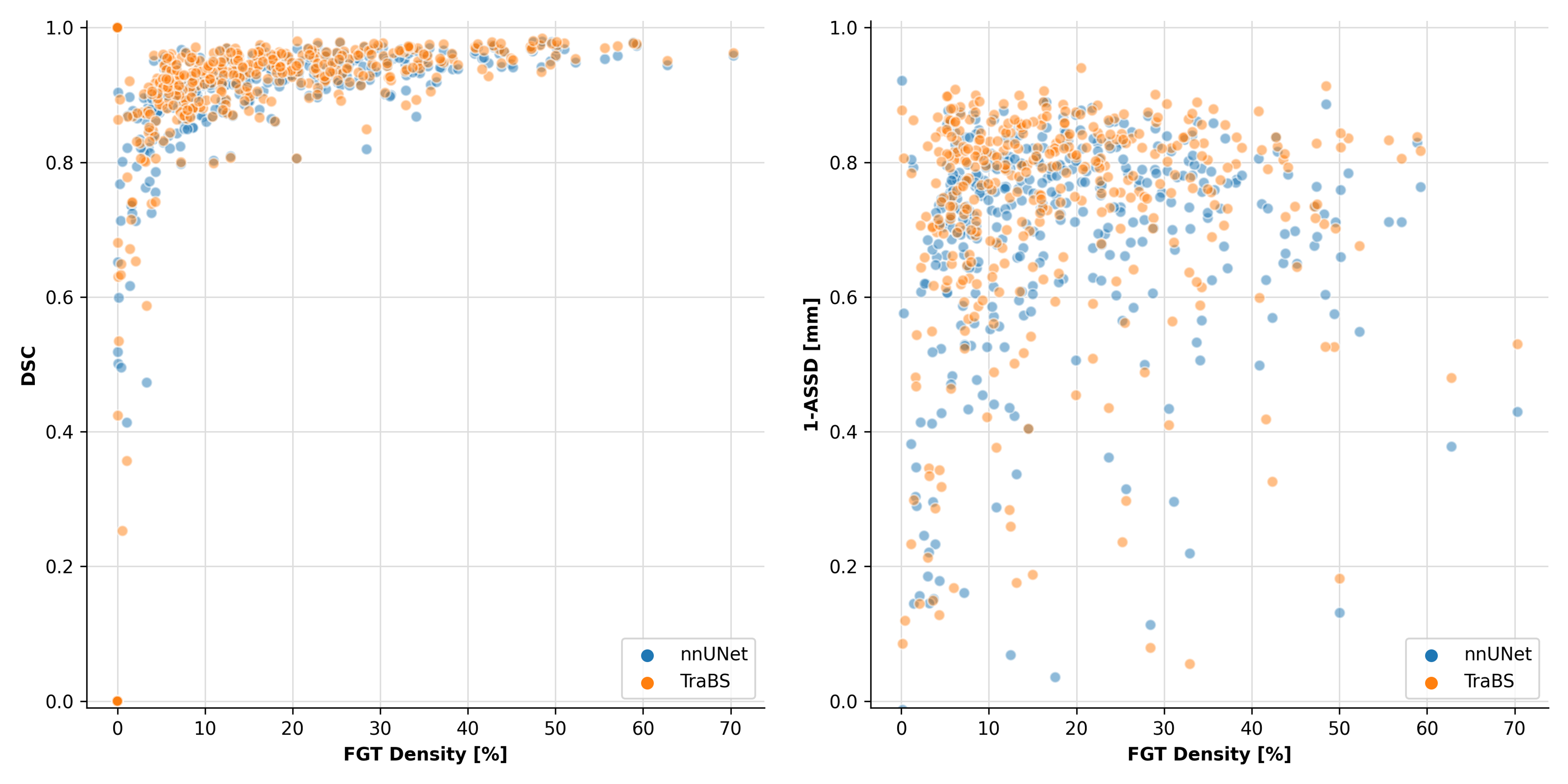}
    \caption{Dice Similarity Coefficient (DSC) and Average Symmetric Surface Distance (ASSD) between the automated and manual segmentations for all examined neural network architectures. Independent of the neural network used, DSC was lower in examinations of low-density breasts, while ASSD was not influenced by breast density.}
    \label{fig:3}
\end{figure}

We tested the segmentation performance of TraBS as compared to nnUNet in terms of overlap between the ground truth and the automated segmentation. nnUNet achieved a mean Dice score of 0.909±0.069 for the FGT segmentation on our internal dataset, see Table \ref{tab:3}. Our improved model TraBS achieved consistently better results with a mean Dice score of 0.916±0.067, P<0.001. TraBS also demonstrated a lower Average Symmetric Surface Distance (ASSD) (0.548±2.195) than nnUNet (0.657±2.856, P=0.001), indicating, that finer details are more accurately assessed by TraBS. By trend, segmentation performance as measured by Dice score was lower for both models when breasts were less dense, i.e., when the fractional volume of FGT within the breast was lower, see Figure \ref{fig:3}. 

\begin{table}[]
\caption{Comparison of the Dice Score (DSC, higher is better), Average Symmetric Surface Distance (ASSD, lower is better), Pearson correlation coefficients between quantitative estimation by radiologists and neural network of the Breast Density ($\rho_{Dense}$, higher is better) and background parenchymal enhancement ($\rho_{BPE}$, higher is better) on the UKA dataset. The best performance is denoted in bold.}
\label{tab:3}
\centering
\begin{tabular}{@{}lllll@{}}
\toprule
             & DSC                  & ASSD [mm]            & $\rho_{Dense}$ & $\rho_{BPE}$          \\
\midrule
nnUNet       & 0.909±0.069          & 0.657±2.856          & 0.995          & 0.950          \\
TraBS (ours) & \textbf{0.916±0.067} & \textbf{0.548±2.195} & \textbf{0.996} & \textbf{0.992} \\
\bottomrule
\end{tabular}
\end{table}

\subsubsection*{Fine Details and overall Structure are better captured by TraBS}
\begin{figure}
    \centering
    \includegraphics[width=\textwidth]{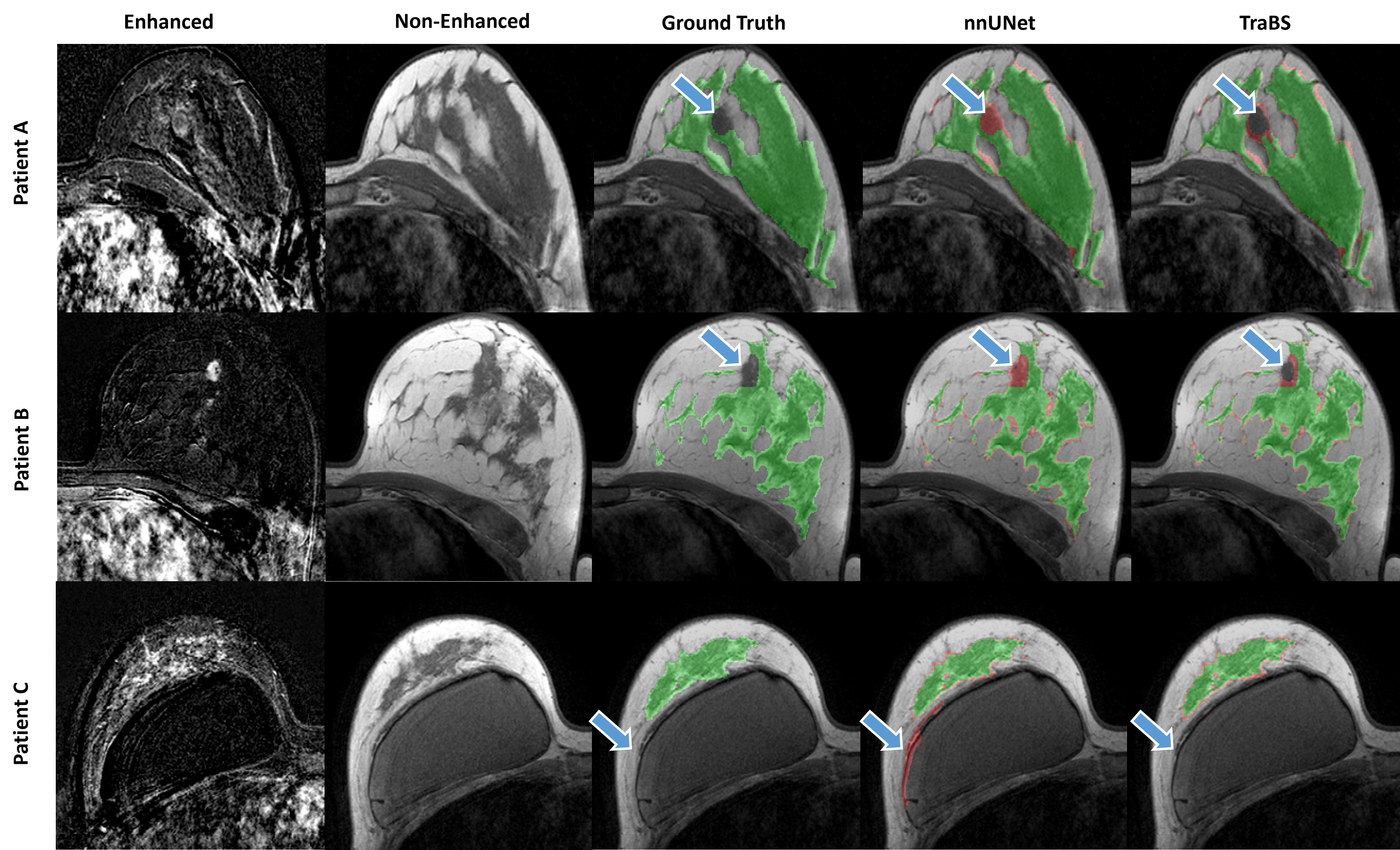}
    \caption{Sample MRI examinations of the internal UKA dataset. The two leftmost columns show the contrast-enhanced subtraction and non-enhanced T1-weighted image. The third column shows the ground truth segmentation by the radiologists and the two remaining columns show the segmentations by the neural networks. Correct segmentations are displayed in green and incorrectly labeled regions in red. Blue arrows denote challenging regions such as lesions (Patient A and B) or breast implants (Patient C).}
    \label{fig:4}
\end{figure}

In addition to quantitative assessment, an expert radiologist visually assessed the segmentation quality and found that TraBS performed better in capturing both overall structure and fine details compared to nnUNet. Specifically, TraBS was better at differentiating between breast implants and FGT and distinguishing between lesions and normal breast tissue, as shown in Figure \ref{fig:4}.

\subsubsection*{Transformer-based segmentation translates to more accurate Clinical Measures}
To investigate how the segmentation performance relates to clinical measurements used to assess patients’ risks such as breast density, we examined the correlation between such measures when calculated based on the ground truth segmentation and on the automated segmentation. Both nnUNet and TraBS demonstrated an almost perfect correlation to the manually derived breast density and BPE, as shown in Table \ref{tab:3}. Although the correlations were almost perfect for both models, TraBS showed slightly higher correlations (P=0.11 and P=0.06).

\subsection{Transformer outperforms CNNs on the External Datasets}
\subsubsection*{Overall segmentation performance}
Applying the models to unseen external datasets with differing MRI sequence protocols resulted in overall lower performance (Table \ref{tab:4}).
However, TraBS still performed better than nnUNet, achieving a mean Dice score of 0.864±0.081 for the DUKE dataset, compared to nnUNet's score of 0.824±0.144 (P=0.004).  Similarly, ASSD was higher (=worse) for nnUNet (0.727±0.620) than for TraBS (0.584±0.413, P=0.034).

\begin{table}[]
\caption{Comparison of the Dice Score (DSC, higher is better), Average Symmetric Surface Distance (ASSD, lower is better), Pearson correlation coefficients between quantitative estimation by radiologists and neural network of the Breast Density ($\rho_{Dense}$, higher is better) and background parenchymal enhancement ($\rho_{BPE}$, higher is better) in the external DUKE dataset. The best performance is denoted in bold. }
\label{tab:4}
\centering
\begin{tabular}{@{}lllll@{}}
\toprule
             & DSC                  & ASSD [mm]            & $\rho_{Dense}$ & $\rho_{BPE}$            \\
\midrule
nnUNet       & 0.824±0.144          & 0.727±0.620          & 0.901          & 0.979          \\
TraBS (ours) & \textbf{0.864±0.081} & \textbf{0.584±0.413} & \textbf{0.955} & \textbf{0.987} \\
\bottomrule
\end{tabular}
\end{table}

\subsubsection*{Fine Details and overall Structure are better captured by TraBS}
\begin{figure}
    \centering
    \includegraphics[width=\textwidth]{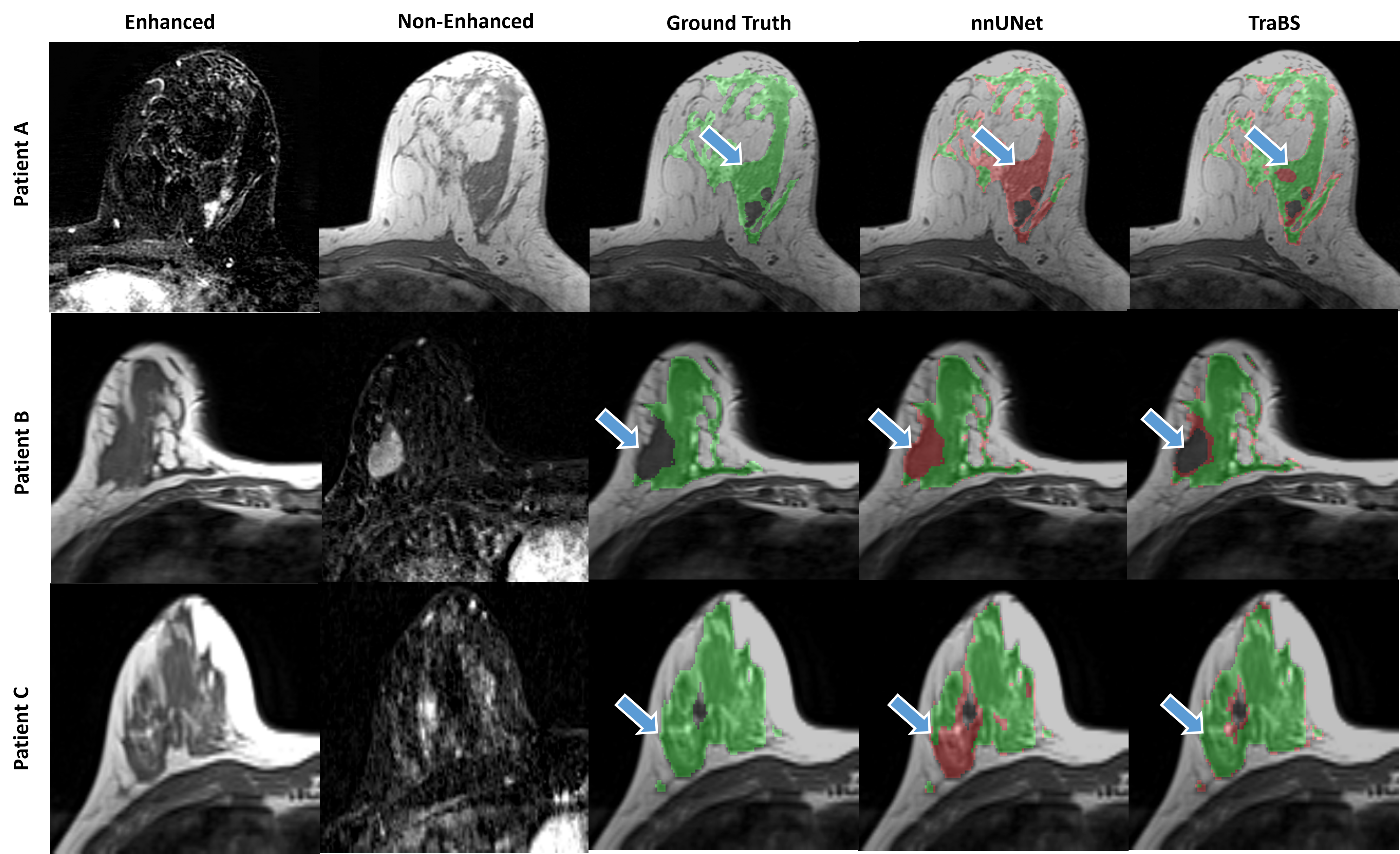}
    \caption{Sample MRI examinations of the external DUKE dataset. The two leftmost columns show the contrast-enhanced subtraction and non-enhanced T1-weighted image. The third column shows the ground truth segmentation by the radiologists and the two remaining columns show the segmentations by the neural networks. Correct segmentations are displayed in green and incorrectly labeled regions in red. Blue arrows denote challenging regions such as lesions (Patient A and B) or breast implants (Patient C).}
    \label{fig:5}
\end{figure}

Visual inspection of the segmentations in the external datasets confirmed the superior performance of TraBS, as it was better able to capture fine details and overall structure compared to nnUNet. Sample images are given in Figure \ref{fig:5}.

\subsubsection*{Transformer-based segmentation translates to more accurate Clinical Measures}
Despite the limited overall segmentation quality on the external DUKE dataset, both TraBS and nnUNet still demonstrated good correlations with manual segmentations for breast density and BPE (Table \ref{tab:4}). However, TraBS achieved higher correlations than nnUNet (P=0.007 and P=0.24).

\section{Discussion}

In this study, we propose a novel network architecture, TraBS, for segmenting fibroglandular tissue (FGT) in breast MRI images. We demonstrate that TraBS outperforms the previous state of the art in both internal and external validation sets. Breast density and BPE are important factors in determining patients’ cancer risk. Thus, accurate and reliable methods for the automated extraction of quantitative markers such as breast density and BPE are needed. Our work advances the field in four aspects:
First, all groups who have applied neural networks on FGT segmentation have only evaluated their algorithms on internal test sets, i.e., examinations that are similar in appearance to the examinations upon which the algorithm was trained, see Table \ref{tab:2} for an overview of previous research. This is a shortcoming that needs to be addressed in view of the plethora of MRI scanner protocols that are currently in clinical use. We tackled this gap by evaluating our proposed TraBS model on an external dataset and we demonstrated that the new transformer-based architecture exhibits a better generalization performance as compared to nnUNet.
Second, we examined the Dice score as a function of breast density and found that lower FGT density results in a lower Dice score. This partly explains the spread of reported Dice scores in the literature (Table \ref{tab:2}), as the test set that is used for evaluation has a large effect on the Dice metric: if segmentation algorithms are tested on breast MRI examinations with high amounts of FGT, Dice scores are higher by trend. This is an important finding for future studies and therefore, we suggest that future works on FGT segmentation should contain a report about the mean FGT density of the test set or a graph similar to Figure \ref{fig:3}.
Third, we make the manual segmentations for the DUKE data publicly available to serve as a reference standard for future evaluations. We reckon that this can contribute to independent external evaluations of segmentation algorithms for breast MRI.
Last but not least, we demonstrate the overall better performance of our transformer-based model TraBS as compared to the previous state-of-the-art architecture for breast tissue segmentation in all selected performance metrics. We make our code publicly available, alongside the trained model, to further advance the field and to bridge the gap to clinical application.
Our work has limitations that relate to the fact that manual segmentations are extremely time-consuming to obtain: First, even though we evaluated the model on external test data, we did not include any external training data. Thus, the segmentation performance decreases when applied to external data and even though TraBS is more robust to domain shift, its performance could be increased by including additional multi-domain data during training. Future work should focus on this in order to make the segmentation performance more robust. Second, we only included 40 external examinations as test cases from one external institution. Even though this is progress as compared to previous research, the database for a broad multi-institutional can and should be extended to provide a global perspective on the performance in so far underrepresented patient groups. Third, we did not investigate inter-rater variability due to the lack of multiple segmentations by multiple readers on the same examinations. This should be done by future studies in order to evaluate the accuracy of the segmentations which so far serve as ground truth, i.e., the human-generated segmentations.

\begin{table}[]
\caption{Comparison of studies using neural networks for fibroglandular tissue (FGT) segmentation in MRI. The number of patients refers to the test set(s). The agreement between the manual and neural network segmentation was measured based on the Dice Similarity Coefficient (DSC), the Pearson correlation coefficient ($\rho_{FGT}$) of the FGT volume, the breast density ($\rho_{Dense}$ ), and the correlation ($\rho_{BPE}$ ) of the background parenchymal enhancement (BPE). Note: *BPE was estimated qualitatively by radiologists, **Spearman Correlation Coefficient}
\centering
\label{tab:2}
\begin{tabular}{@{}lllllll@{}}
\toprule
Study                          & Patients & Region      & DSC & $\rho_{FGT}$ & $\rho_{Dense}$& $\rho_{BPE}$   \\ \midrule
Ivanovska \cite{ivanovska2019} & 40       & Germany     & 0.93±0.01 & -         & -      & -     \\
Huo \cite{huo2021}             & 100      & China       & 0.88±0.08 & 0.972     & 0.981  & -     \\
Ma \cite{ma2020}               & 100      & -           & 0.87±0.07 & -         & -      & 0.63* \\
Dalmis \cite{dalmis2017}       & 66       & Netherlands & 0.85±0.09 & -         & 0.974  & -     \\
Nam \cite{nam2021}             & 200      & Korea       & 0.85±0.11 & 0.93**    & -      & -     \\
Zhang \cite{zhang2019}         & 28       & Asia        & 0.83±0.06 & 0.97-0.99 & -      & -     \\
Zhang \cite{zhang2021}         & 40       & -           & 0.81±0.08 & 0.956     & -      & -     \\
Ha \cite{ha2019}               & 137      & USA         & 0.81      & 0.998     & -      & 0.955 \\
Ours &
  \begin{tabular}[c]{@{}l@{}}200, \\  40\end{tabular} &
  \begin{tabular}[c]{@{}l@{}}Germany, \\ USA\end{tabular} &
  \begin{tabular}[c]{@{}l@{}}0.92±0.07\\ 0.86±0.08\end{tabular} &
  \begin{tabular}[c]{@{}l@{}}0.996\\ 0.945\end{tabular} &
  \begin{tabular}[c]{@{}l@{}}0.996\\ 0.955\end{tabular} &
  \begin{tabular}[c]{@{}l@{}}0.992\\ 0.987\end{tabular} \\ \bottomrule
\end{tabular}
\end{table}

\section{Conclusion}
In conclusion, our proposed TraBS network demonstrates excellent performance in segmenting FGT in breast MRI images. This paves the way for routine automated FGT segmentation and automatic quantification of breast density and BPE.

\section*{Conflicts of Interest}
The authors declare that they have no competing interests.

\bibliographystyle{unsrtnat}
\bibliography{references}  %%% Uncomment this line and comment out the ``thebibliography'' section below to use the external .bib file (using bibtex) .

\begin{thebibliography}{30}
\providecommand{\natexlab}[1]{#1}
\providecommand{\url}[1]{\texttt{#1}}
\expandafter\ifx\csname urlstyle\endcsname\relax
  \providecommand{\doi}[1]{doi: #1}\else
  \providecommand{\doi}{doi: \begingroup \urlstyle{rm}\Url}\fi

\bibitem[Siegel et~al.(2022)Siegel, Miller, Fuchs, and Jemal]{Siegel2022-kj}
Rebecca~L Siegel, Kimberly~D Miller, Hannah~E Fuchs, and Ahmedin Jemal.
\newblock Cancer statistics, 2022.
\newblock \emph{CA Cancer J. Clin.}, 72\penalty0 (1):\penalty0 7--33, January
  2022.

\bibitem[Mann et~al.(2022)Mann, Athanasiou, Baltzer, {Camps-Herrero}, Clauser,
  Fallenberg, Forrai, Fuchsj{\"a}ger, Helbich, {Killburn-Toppin}, Lesaru,
  Panizza, Pediconi, Pijnappel, Pinker, Sardanelli, Sella, {Thomassin-Naggara},
  Zackrisson, Gilbert, Kuhl, and {On behalf of the European Society of Breast
  Imaging (EUSOBI)}]{mann2022}
Ritse~M. Mann, Alexandra Athanasiou, Pascal A.~T. Baltzer, Julia
  {Camps-Herrero}, Paola Clauser, Eva~M. Fallenberg, Gabor Forrai, Michael~H.
  Fuchsj{\"a}ger, Thomas~H. Helbich, Fleur {Killburn-Toppin}, Mihai Lesaru,
  Pietro Panizza, Federica Pediconi, Ruud~M. Pijnappel, Katja Pinker, Francesco
  Sardanelli, Tamar Sella, Isabelle {Thomassin-Naggara}, Sophia Zackrisson,
  Fiona~J. Gilbert, Christiane~K. Kuhl, and {On behalf of the European Society
  of Breast Imaging (EUSOBI)}.
\newblock Breast cancer screening in women with extremely dense breasts
  recommendations of the {{European Society}} of {{Breast Imaging}}
  ({{EUSOBI}}).
\newblock \emph{European Radiology}, 32\penalty0 (6):\penalty0 4036--4045, June
  2022.
\newblock ISSN 1432-1084.
\newblock \doi{10.1007/s00330-022-08617-6}.

\bibitem[Oeffinger et~al.(2015)Oeffinger, Fontham, Etzioni, Herzig, Michaelson,
  Shih, Walter, Church, Flowers, LaMonte, Wolf, DeSantis, {Lortet-Tieulent},
  Andrews, {Manassaram-Baptiste}, Saslow, Smith, Brawley, Wender, and {American
  Cancer Society}]{Oeffinger2015-yz}
Kevin~C Oeffinger, Elizabeth T~H Fontham, Ruth Etzioni, Abbe Herzig, James~S
  Michaelson, Ya-Chen~Tina Shih, Louise~C Walter, Timothy~R Church,
  Christopher~R Flowers, Samuel~J LaMonte, Andrew M~D Wolf, Carol DeSantis,
  Joannie {Lortet-Tieulent}, Kimberly Andrews, Deana {Manassaram-Baptiste},
  Debbie Saslow, Robert~A Smith, Otis~W Brawley, Richard Wender, and {American
  Cancer Society}.
\newblock Breast cancer screening for women at average risk: 2015 guideline
  update from the american cancer society.
\newblock \emph{JAMA : the journal of the American Medical Association},
  314\penalty0 (15):\penalty0 1599--1614, October 2015.

\bibitem[Magni et~al.(2022)Magni, Interlenghi, Cozzi, Al{\`i}, Salvatore,
  Azzena, Capra, Carriero, Della~Pepa, Fazzini, Granata, Monti, Muscogiuri,
  Pellegrino, Schiaffino, Castiglioni, Papa, and Sardanelli]{Magni2022-av}
Veronica Magni, Matteo Interlenghi, Andrea Cozzi, Marco Al{\`i}, Christian
  Salvatore, Alcide~A Azzena, Davide Capra, Serena Carriero, Gianmarco
  Della~Pepa, Deborah Fazzini, Giuseppe Granata, Caterina~B Monti, Giulia
  Muscogiuri, Giuseppe Pellegrino, Simone Schiaffino, Isabella Castiglioni,
  Sergio Papa, and Francesco Sardanelli.
\newblock Development and validation of an {{AI-driven}} mammographic breast
  density classification tool based on radiologist consensus.
\newblock \emph{Radiology. Artificial intelligence}, 4\penalty0 (2):\penalty0
  e210199, March 2022.

\bibitem[Sindi et~al.(2019)Sindi, S{\'a}~Dos~Reis, Bennett, Stevenson, and
  Sun]{sindi2019}
Rooa Sindi, Cl{\'a}udia S{\'a}~Dos~Reis, Colleen Bennett, Gil Stevenson, and
  Zhonghua Sun.
\newblock Quantitative {{Measurements}} of {{Breast Density Using Magnetic
  Resonance Imaging}}: {{A Systematic Review}} and {{Meta-Analysis}}.
\newblock \emph{Journal of Clinical Medicine}, 8\penalty0 (5):\penalty0 745,
  May 2019.
\newblock ISSN 2077-0383.
\newblock \doi{10.3390/jcm8050745}.

\bibitem[Thompson et~al.(2019)Thompson, Mallawaarachchi, Dwivedi, Ayyappan,
  Shokar, Lakshmanaswamy, and Dwivedi]{thompson2019}
Christopher~M. Thompson, Indika Mallawaarachchi, Durgesh~K. Dwivedi, Anoop~P.
  Ayyappan, Navkiran~K. Shokar, Rajkumar Lakshmanaswamy, and Alok~K. Dwivedi.
\newblock The {{Association}} of {{Background Parenchymal Enhancement}} at
  {{Breast MRI}} with {{Breast Cancer}}: {{A Systematic Review}} and
  {{Meta-Analysis}}.
\newblock \emph{Radiology}, 292\penalty0 (3):\penalty0 552--561, September
  2019.
\newblock ISSN 0033-8419, 1527-1315.
\newblock \doi{10.1148/radiol.2019182441}.

\bibitem[Dalm{\i}{\c s} et~al.(2017)Dalm{\i}{\c s}, Litjens, Holland, Setio,
  Mann, Karssemeijer, and {Gubern-M{\'e}rida}]{dalmis2017}
Mehmet~Ufuk Dalm{\i}{\c s}, Geert Litjens, Katharina Holland, Arnaud Setio,
  Ritse Mann, Nico Karssemeijer, and Albert {Gubern-M{\'e}rida}.
\newblock Using deep learning to segment breast and fibroglandular tissue in
  {{MRI}} volumes.
\newblock \emph{Medical Physics}, 44\penalty0 (2):\penalty0 533--546, February
  2017.
\newblock ISSN 00942405.
\newblock \doi{10.1002/mp.12079}.

\bibitem[Huo et~al.(2021)Huo, Hu, Xiao, Gu, Chu, and Jiang]{huo2021}
Lu~Huo, Xiaoxin Hu, Qin Xiao, Yajia Gu, Xu~Chu, and Luan Jiang.
\newblock Segmentation of whole breast and fibroglandular tissue using
  {{nnU-Net}} in dynamic contrast enhanced {{MR}} images.
\newblock \emph{Magnetic Resonance Imaging}, 82:\penalty0 31--41, October 2021.
\newblock ISSN 0730725X.
\newblock \doi{10.1016/j.mri.2021.06.017}.

\bibitem[Zhang et~al.(2019)Zhang, Chen, Chang, Park, Kim, Chan, Chang, Chow,
  Luk, Kwong, and Su]{zhang2019}
Yang Zhang, Jeon-Hor Chen, Kai-Ting Chang, Vivian~Youngjean Park, Min~Jung Kim,
  Siwa Chan, Peter Chang, Daniel Chow, Alex Luk, Tiffany Kwong, and Min-Ying
  Su.
\newblock Automatic {{Breast}} and {{Fibroglandular Tissue Segmentation}} in
  {{Breast MRI Using Deep Learning}} by a {{Fully-Convolutional Residual Neural
  Network U-Net}}.
\newblock \emph{Academic Radiology}, 26\penalty0 (11):\penalty0 1526--1535,
  November 2019.
\newblock ISSN 10766332.
\newblock \doi{10.1016/j.acra.2019.01.012}.

\bibitem[Ghaffari~Laleh et~al.(2022)Ghaffari~Laleh, Truhn, Veldhuizen, Han,
  {van Treeck}, Buelow, Langer, Dislich, Boor, Schulz, and
  Kather]{ghaffarilaleh2022a}
Narmin Ghaffari~Laleh, Daniel Truhn, Gregory~Patrick Veldhuizen, Tianyu Han,
  Marko {van Treeck}, Roman~D. Buelow, Rupert Langer, Bastian Dislich, Peter
  Boor, Volkmar Schulz, and Jakob~Nikolas Kather.
\newblock Adversarial attacks and adversarial robustness in computational
  pathology.
\newblock \emph{Nature Communications}, 13\penalty0 (1):\penalty0 5711,
  September 2022.
\newblock ISSN 2041-1723.
\newblock \doi{10.1038/s41467-022-33266-0}.

\bibitem[Tang et~al.(2022)Tang, Yang, Li, Roth, Landman, Xu, Nath, and
  Hatamizadeh]{tang2022}
Yucheng Tang, Dong Yang, Wenqi Li, Holger~R. Roth, Bennett Landman, Daguang Xu,
  Vishwesh Nath, and Ali Hatamizadeh.
\newblock Self-{{Supervised Pre-Training}} of {{Swin Transformers}} for {{3D
  Medical Image Analysis}}.
\newblock In \emph{2022 {{IEEE}}/{{CVF Conference}} on {{Computer Vision}} and
  {{Pattern Recognition}} ({{CVPR}})}, number arXiv:2111.14791, pages
  20698--20708, {New Orleans, LA, USA}, June 2022. {IEEE}.
\newblock ISBN 978-1-66546-946-3.
\newblock \doi{10.1109/CVPR52688.2022.02007}.

\bibitem[Devlin et~al.(2019)Devlin, Chang, Lee, and Toutanova]{devlin2019}
Jacob Devlin, Ming-Wei Chang, Kenton Lee, and Kristina Toutanova.
\newblock {{BERT}}: {{Pre-training}} of {{Deep Bidirectional Transformers}} for
  {{Language Understanding}}.
\newblock \emph{arXiv:1810.04805 [cs]}, May 2019.
\newblock \doi{10.48550/arXiv.1810.04805}.

\bibitem[Radford et~al.()Radford, Narasimhan, Salimans, and
  Sutskever]{radforda}
Alec Radford, Karthik Narasimhan, Tim Salimans, and Ilya Sutskever.
\newblock Improving {{Language Understanding}} by {{Generative Pre-Training}}.

\bibitem[Saha et~al.(2022)Saha, Harowicz, Grimm, Weng, Cain, Kim, Ghate, Walsh,
  and Mazurowski]{saha2022}
Ashirbani Saha, Michael~R. Harowicz, Lars~J. Grimm, Jingxi Weng, E.~H. Cain,
  C.~E. Kim, S.~V. Ghate, R.~Walsh, and Maciej~A. Mazurowski.
\newblock Dynamic contrast-enhanced magnetic resonance images of breast cancer
  patients with tumor locations, August 2022.

\bibitem[{M{\"u}ller-Franzes} et~al.(2023){M{\"u}ller-Franzes}, Huck,
  Tayebi~Arasteh, Khader, Han, Schulz, Dethlefsen, Kather, Nebelung, Nolte,
  Kuhl, and Truhn]{muller-franzes2023}
Gustav {M{\"u}ller-Franzes}, Luisa Huck, Soroosh Tayebi~Arasteh, Firas Khader,
  Tianyu Han, Volkmar Schulz, Ebba Dethlefsen, Jakob~Nikolas Kather, Sven
  Nebelung, Teresa Nolte, Christiane Kuhl, and Daniel Truhn.
\newblock Using {{Machine Learning}} to {{Reduce}} the {{Need}} for {{Contrast
  Agents}} in {{Breast MRI}} through {{Synthetic Images}}.
\newblock \emph{Radiology}, page 222211, March 2023.
\newblock ISSN 0033-8419, 1527-1315.
\newblock \doi{10.1148/radiol.222211}.

\bibitem[Kuhl et~al.(2017)Kuhl, Strobel, Bieling, Leutner, Schild, and
  Schrading]{kuhl2017}
Christiane~K. Kuhl, Kevin Strobel, Heribert Bieling, Claudia Leutner, Hans~H.
  Schild, and Simone Schrading.
\newblock Supplemental {{Breast MR Imaging Screening}} of {{Women}} with
  {{Average Risk}} of {{Breast Cancer}}.
\newblock \emph{Radiology}, 283\penalty0 (2):\penalty0 361--370, May 2017.
\newblock ISSN 0033-8419, 1527-1315.
\newblock \doi{10.1148/radiol.2016161444}.

\bibitem[Yushkevich et~al.(2006)Yushkevich, Piven, Hazlett, Smith, Ho, Gee, and
  Gerig]{yushkevich2006}
Paul~A. Yushkevich, Joseph Piven, Heather~Cody Hazlett, Rachel~Gimpel Smith,
  Sean Ho, James~C. Gee, and Guido Gerig.
\newblock User-guided {{3D}} active contour segmentation of anatomical
  structures: {{Significantly}} improved efficiency and reliability.
\newblock \emph{NeuroImage}, 31\penalty0 (3):\penalty0 1116--1128, July 2006.
\newblock ISSN 10538119.
\newblock \doi{10.1016/j.neuroimage.2006.01.015}.

\bibitem[Paszke et~al.(2019)Paszke, Gross, Massa, Lerer, Bradbury, Chanan,
  Killeen, Lin, Gimelshein, Antiga, Desmaison, K{\"o}pf, Yang, DeVito, Raison,
  Tejani, Chilamkurthy, Steiner, Fang, Bai, and Chintala]{paszke2019}
Adam Paszke, Sam Gross, Francisco Massa, Adam Lerer, James Bradbury, Gregory
  Chanan, Trevor Killeen, Zeming Lin, Natalia Gimelshein, Luca Antiga, Alban
  Desmaison, Andreas K{\"o}pf, Edward Yang, Zach DeVito, Martin Raison, Alykhan
  Tejani, Sasank Chilamkurthy, Benoit Steiner, Lu~Fang, Junjie Bai, and Soumith
  Chintala.
\newblock {{PyTorch}}: {{An Imperative Style}}, {{High-Performance Deep
  Learning Library}}.
\newblock 2019.
\newblock \doi{10.48550/ARXIV.1912.01703}.

\bibitem[Hatamizadeh et~al.(2022)Hatamizadeh, Nath, Tang, Yang, Roth, and
  Xu]{hatamizadeh2022}
Ali Hatamizadeh, Vishwesh Nath, Yucheng Tang, Dong Yang, Holger Roth, and
  Daguang Xu.
\newblock Swin {{UNETR}}: {{Swin Transformers}} for {{Semantic Segmentation}}
  of {{Brain Tumors}} in {{MRI Images}}.
\newblock {arXiv}, 2022.
\newblock \doi{10.1007/978-3-031-08999-2_22}.

\bibitem[Isensee et~al.(2021)Isensee, Jaeger, Kohl, Petersen, and
  {Maier-Hein}]{isensee2021}
Fabian Isensee, Paul~F. Jaeger, Simon A.~A. Kohl, Jens Petersen, and Klaus~H.
  {Maier-Hein}.
\newblock {{nnU-Net}}: A self-configuring method for deep learning-based
  biomedical image segmentation.
\newblock \emph{Nature Methods}, 18\penalty0 (2):\penalty0 203--211, February
  2021.
\newblock ISSN 1548-7091, 1548-7105.
\newblock \doi{10.1038/s41592-020-01008-z}.

\bibitem[Ma et~al.(2021)Ma, Chen, Ng, Huang, Li, Li, Yang, and Martel]{ma2021}
Jun Ma, Jianan Chen, Matthew Ng, Rui Huang, Yu~Li, Chen Li, Xiaoping Yang, and
  Anne~L. Martel.
\newblock Loss odyssey in medical image segmentation.
\newblock \emph{Medical Image Analysis}, 71:\penalty0 102035, July 2021.
\newblock ISSN 13618415.
\newblock \doi{10.1016/j.media.2021.102035}.

\bibitem[{P{\'e}rez-Garc{\'i}a} et~al.(2020){P{\'e}rez-Garc{\'i}a}, Sparks, and
  Ourselin]{perez-garcia2020}
Fernando {P{\'e}rez-Garc{\'i}a}, Rachel Sparks, and S{\'e}bastien Ourselin.
\newblock {{TorchIO}}: {{A Python}} library for efficient loading,
  preprocessing, augmentation and patch-based sampling of medical images in
  deep learning.
\newblock 2020.
\newblock \doi{10.48550/ARXIV.2003.04696}.

\bibitem[Zou et~al.(2004)Zou, Warfield, Bharatha, Tempany, Kaus, Haker, Wells,
  Jolesz, and Kikinis]{zou2004}
Kelly~H. Zou, Simon~K. Warfield, Aditya Bharatha, Clare~M.C. Tempany,
  Michael~R. Kaus, Steven~J. Haker, William~M. Wells, Ferenc~A. Jolesz, and Ron
  Kikinis.
\newblock Statistical validation of image segmentation quality based on a
  spatial overlap index1.
\newblock \emph{Academic Radiology}, 11\penalty0 (2):\penalty0 178--189,
  February 2004.
\newblock ISSN 10766332.
\newblock \doi{10.1016/S1076-6332(03)00671-8}.

\bibitem[Nai et~al.(2021)Nai, Teo, Tan, O'Doherty, Stephenson, Thian, Chiong,
  and Reilhac]{Nai2021-wo}
Ying-Hwey Nai, Bernice~W Teo, Nadya~L Tan, Sophie O'Doherty, Mary~C Stephenson,
  Yee~Liang Thian, Edmund Chiong, and Anthonin Reilhac.
\newblock Comparison of metrics for the evaluation of medical segmentations
  using prostate {{MRI}} dataset.
\newblock \emph{Computers in Biology and Medicine}, 134:\penalty0 104497, July
  2021.

\bibitem[Amrhein et~al.(2019)Amrhein, Greenland, and McShane]{Amrhein2019-wr}
Valentin Amrhein, Sander Greenland, and Blake McShane.
\newblock Scientists rise up against statistical significance.
\newblock \emph{Nature}, 567\penalty0 (7748):\penalty0 305--307, March 2019.

\bibitem[Ivanovska et~al.(2019)Ivanovska, Jentschke, Daboul, Hegenscheid,
  V{\"o}lzke, and W{\"o}rg{\"o}tter]{ivanovska2019}
Tatyana Ivanovska, Thomas~G. Jentschke, Amro Daboul, Katrin Hegenscheid, Henry
  V{\"o}lzke, and Florentin W{\"o}rg{\"o}tter.
\newblock A deep learning framework for efficient analysis of breast volume and
  fibroglandular tissue using {{MR}} data with strong artifacts.
\newblock \emph{International Journal of Computer Assisted Radiology and
  Surgery}, 14\penalty0 (10):\penalty0 1627--1633, October 2019.
\newblock ISSN 1861-6410, 1861-6429.
\newblock \doi{10.1007/s11548-019-01928-y}.

\bibitem[Ma et~al.(2020)Ma, Wang, Zheng, Liu, Long, Zhang, Wei, and Lu]{ma2020}
Xiangyuan Ma, Jinlong Wang, Xinpeng Zheng, Zhuangsheng Liu, Wansheng Long,
  Yaqin Zhang, Jun Wei, and Yao Lu.
\newblock Automated fibroglandular tissue segmentation in breast {{MRI}} using
  generative adversarial networks.
\newblock \emph{Physics in Medicine \& Biology}, 65\penalty0 (10):\penalty0
  105006, May 2020.
\newblock ISSN 1361-6560.
\newblock \doi{10.1088/1361-6560/ab7e7f}.

\bibitem[Nam et~al.(2021)Nam, Park, Kang, and Kim]{nam2021}
Yoonho Nam, Ga~Eun Park, Junghwa Kang, and Sung~Hun Kim.
\newblock Fully {{Automatic Assessment}} of {{Background Parenchymal
  Enhancement}} on {{Breast MRI Using Machine}}-{{Learning Models}}.
\newblock \emph{Journal of Magnetic Resonance Imaging}, 53\penalty0
  (3):\penalty0 818--826, March 2021.
\newblock ISSN 1053-1807, 1522-2586.
\newblock \doi{10.1002/jmri.27429}.

\bibitem[Zhang et~al.(2021)Zhang, Chan, Chen, Chang, Lin, Pan, Lin, Kwong,
  Parajuli, Mehta, Chien, and Su]{zhang2021}
Yang Zhang, Siwa Chan, Jeon-Hor Chen, Kai-Ting Chang, Chin-Yao Lin, Huay-Ben
  Pan, Wei-Ching Lin, Tiffany Kwong, Ritesh Parajuli, Rita~S. Mehta, Sou-Hsin
  Chien, and Min-Ying Su.
\newblock Development of {{U-Net Breast Density Segmentation Method}} for
  {{Fat-Sat MR Images Using Transfer Learning Based}} on {{Non-Fat-Sat Model}}.
\newblock \emph{Journal of Digital Imaging}, 34\penalty0 (4):\penalty0
  877--887, August 2021.
\newblock ISSN 0897-1889, 1618-727X.
\newblock \doi{10.1007/s10278-021-00472-z}.

\bibitem[Ha et~al.(2019)Ha, Chang, Mema, Mutasa, Karcich, Wynn, Liu, and
  Jambawalikar]{ha2019}
Richard Ha, Peter Chang, Eralda Mema, Simukayi Mutasa, Jenika Karcich, Ralph~T.
  Wynn, Michael~Z. Liu, and Sachin Jambawalikar.
\newblock Fully {{Automated Convolutional Neural Network Method}} for
  {{Quantification}} of {{Breast MRI Fibroglandular Tissue}} and {{Background
  Parenchymal Enhancement}}.
\newblock \emph{Journal of Digital Imaging}, 32\penalty0 (1):\penalty0
  141--147, February 2019.
\newblock ISSN 0897-1889, 1618-727X.
\newblock \doi{10.1007/s10278-018-0114-7}.

\end{thebibliography}





% \end{thebibliography}

\newpage
\section*{Supplemental Material}

\begin{table}[h!]
\caption{Mean DSC values within the five-fold cross-validation. }
\label{tab:S1}
\centering
\begin{tabular}{@{}llllll@{}}
\toprule
             & Fold 1               & Fold 2               & Fold 3               & Fold 4               & Fold 5               \\ \midrule
nnUNet       & 0.887±0.100          & 0.919±0.039          & 0.925±0.032          & 0.904±0.065          & 0.913±0.075          \\
TraBS (ours) & \textbf{0.889±0.104} & \textbf{0.923±0.040} & \textbf{0.934±0.030} & \textbf{0.913±0.052} & \textbf{0.922±0.068} \\ \bottomrule
\end{tabular}
\end{table}

\begin{figure}[h!]
    \centering
    \includegraphics[width=\textwidth]{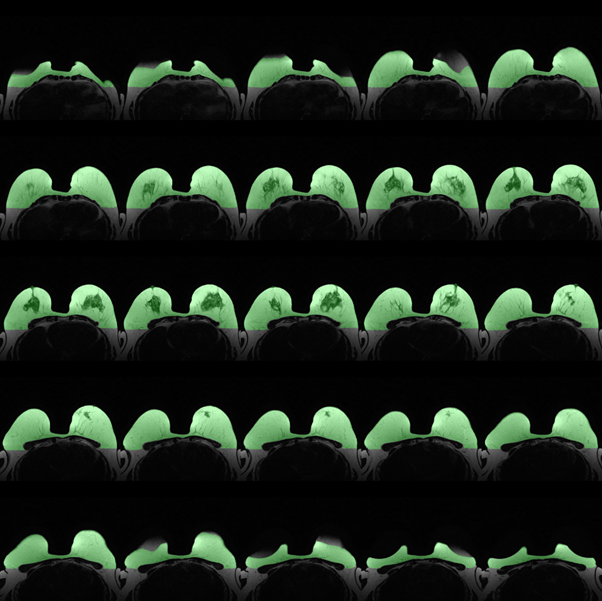}
    \caption{Illustration of the manual breast volume segmentation in the UKA dataset.}
    \label{fig:figS1}
\end{figure}

\begin{figure}[h!]
    \centering
    \includegraphics[width=\textwidth]{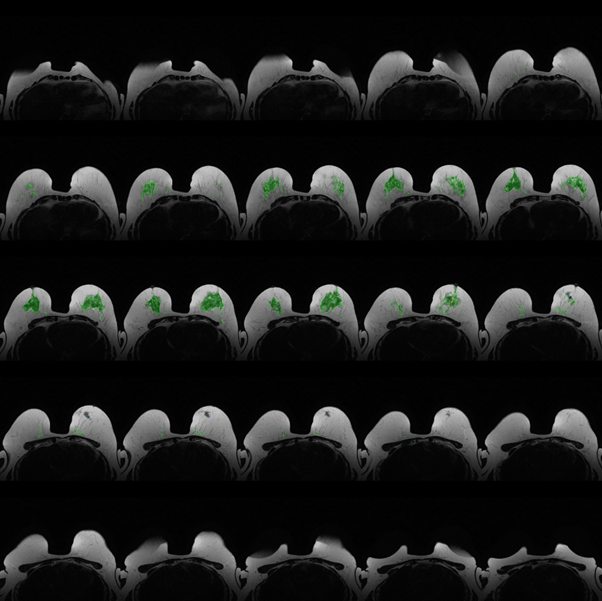}
    \caption{Illustration of the manual fibroglandular tissue segmentation in the UKA dataset. }
    \label{fig:figS2}
\end{figure}

\end{document}